# Superior Resistance Switching in Monolayer MoS$_2$ Channel Based Gated Binary Resistive RAM via Gate-Bias Dependence and a Unique Forming Process

Ansh and Mayank Shrivastava
Department of Electronic Systems Engineering, Indian Institute of Science, Bengaluru-560012, Karnataka, India
Email: mayank@iisc.ac.in

**Abstract**

Resistance switching (RS) in 2-dimensional Molybdenum Disulfide (MoS$_2$) was recently discovered and since then many reports on demonstrating MoS$_2$ RRAM with synapse-like behavior have resulted. These reports strongly justify applications of MoS$_2$ RRAM in neuromorphic hardware as well as an alternative to conventional binary memories. In this work, we unveil the effect of RS, induced by a current-voltage hysteresis cycles across CVD-grown monolayer MoS$_2$ based gated RRAM, on its transistor's electrical and reliability characteristics. A unique gate voltage dependence on the RS is identified which has a remarkable impact on the switching performance of MoS$_2$ RRAM. RS behavior was found to be significantly dependent on the charge conduction in the channel. Moreover, we have shown a potential device forming event when MoS$_2$ gated RRAMs were subjected to a steady-state electrical stress. Both hysteresis and steady-state electrical stress were found to disturb the transistor action of these gated RRAMs, which in fact can be used as a signature of RS. Interestingly, current-voltage hysteresis resulted in unipolar RS, whereas steady-state electrical stress before RS measurement led to bipolar RS. Moreover, successive stress cycles of such an electrical stress leads to multiple resistance states, a behavior similar to synaptic properties like long-term potentiation and long-term depression, typically found in memristors. We find that charge transport mechanism dominant in the MoS$_2$ FET in conjunction with steady-state stress induced device forming determine the extent of RS induced in these MoS$_2$ based gated RRAMs. Finally on the basis of insights developed from the dependence on charge transport mechanism and steady-state stress induced forming of MoS$_2$ channel, we propose a certain steady-state electrical stress condition which can be used as a "forming" process employed prior to use of MoS$_2$ based binary RRAMs for improved switching performance.

## I. Introduction

In order to reduce device footprint to enhance functionality and storage density[1-4], two terminal vertically stacked memory devices like RRAMs etc. have been extensively explored. Trivial device structure along with unique resistance switching mechanism in certain materials justifies their potential for memory applications. Phase change memory (PCM)[5], metal-oxide resistive random-access memory (RRAM)[6], spin-torque transfer RAM (STT-RAM)[7] etc. are various devices that have emerged lately as alternatives for Silicon based flash memory. Interestingly, different mechanisms are responsible for the observed resistance switching (RS) in these devices, for example, joule heating induced crystallization and amorphization in PCM, electric field induced migration of oxygen vacancies resulting in formation of conductive paths in metal oxide RRAMs, spin-selective current injection induced change in spin orientation of a "free-layer" in STT-RAM, form the basis of next generation memory devices[8]. Presently, there is a plethora of bulk materials that exhibit RS which promises further advancements in the field of semiconductor memory. Moreover, recently observed RS in 2D transition metal dichalcogenides (TMDCs)[9-14] has added another dimension to exploration of TMDCs for RS applications. Unlike bulk RS devices, discussed earlier, RS in 2D TMDCs occurs within an atomically thin film. Hence, continued scaling of memory devices along with 3D integration can make further reduction in cost per bit plausible. Sangwan et. al. reported, for the first time, that CVD MoS$_2$ exhibits RS which is mediated by grain boundaries (GB)[9]. Moreover, such a behavior was found to be gate tunable. In a separate work, they demonstrated spike-time dependent plasticity (STDP), short/long term potentiation and depression (STP, STD, LTP and LTD), properties essential for neuromorphic applications[10]. Cheng et. al. reported RS in 1T metallic phase vertical MoS$_2$ devices followed by demonstration of an odd-symmetric memristive device by stacking two 1T phase MoS$_2$ devices back-to-back[11]. Vertical memristors on monolayer TMDCs was reported by Ge et. al. which was the first demonstration of RS at sub-nanometer dimensions[12]. Arnold et. al. also captured LTP in MoS$_2$ transistors via hysteresis engineering[13]. Krishnaprasad et. al. fabricated CVD-grown MoS$_2$/Graphene memristors which exhibit not only basic properties like STP/STD, LTP/LTP and STDP but also linear synaptic weight update thereby making them suitable for use in hardware for unsupervised learning[14]. Another report on ion-based plasticity of MoS$_2$ memristors by Belete et. al. suggests the role of OH$^-$ ions RS observed in MoS$_2$ devices[15]. Besides RS, 2D-material based FG memory cell was demonstrated by Paul et. al. with MoS$_2$ as the transistor channel, Graphene as the FG and hBN as TO, these devices were fabricated on a truly 2D platform and exhibited excellent data retention and endurance[16]. These reports suggest that MoS$_2$ does exhibit RS and other key behavioral traits that justify its importance in binary memory applications and neuromorphic computing hardware implementation.

In order to observe RS in any material, a current-voltage hysteresis curve is typically obtained. Once a material exhibits RS, the two possible states, in case of binary memory devices - LRS and HRS, are revealed from their hysteretic behavior which, however, need to be established via a suitable voltage sweep in a particular direction. In other words, a current-voltage hysteresis operation induces RS in all 2-terminal RRAMs. While every report on RS attempts to elucidate the fundamental mechanism behind RS along with the changes it induces in the material, very few reports discuss the impact of RS on the device behavior from an electrical point-of-view. Moreover, besides hysteresis, other methods to induce RS are unknown. This is precisely what the authors intend to address in this work. We investigate the impact of current-voltage hysteresis-induced RS on the transistor behavior of CVD-grown monolayer MoS$_2$ devices. Further we identify this change in transistor behavior to be similar to a phenomenon observed earlier[17, 18] in which a steady-state DC electrical stress induces negative shift in the threshold voltage (VT) and poor gate control. Hence, it is speculated that such a long-term electrical stress (LTES) also induces RS in MoS$_2$. To materialize the anticipated RS phenomenon, we observe how polarity of a LTES affects the device behavior. It turns out that the channel resistance attains two different states, LRS and HRS, depending upon the polarity which manifests as bipolar RS in our devices. These results elucidate how RS affects the transistor behavior and clearly justify the fact that besides current-voltage hysteresis, a long-term electrical stress is another way to induce RS in CVD-grown monolayer MoS$_2$. Both hysteresis and LTES result in a peculiar transistor behavior which can be used as a signature for RS. Based on gate voltage dependence of RS observed in our experiments, these findings unveil the possibility of increased density of mid-gap states in MoS$_2$ after hysteresis and LTES cycles. Such insights also reveal that RS realized during OFF state operation of MoS$_2$ transistor results in superior switching performance which can be further improved using a high voltage forming. Based on these results, we propose a forming process for MoS$_2$, like in case of other RS materials[19], which can be used to improve the switching window and hence its importance for binary memory applications.

## II. Results and discussion

MoS$_2$ transistors in the back gated configuration, Fig. 1(a), are fabricated using the process discussed in Fig. S1(a) of supplementary information (SI). A ~ 19 cm$^{-1}$ of separation between the two characteristic Raman modes (A$_{1g}$ and E$_{2g}$), as shown in Fig. S1(b) of SI, implies that devices are fabricated on CVD-grown monolayer MoS$_2$. A scanning electron microscopy (SEM) top view image of the as-fabricated transistor is shown in Fig. 1(b). All subsequent measurements are performed inside a vacuum probe station maintained at 10$^{-4}$ torr.

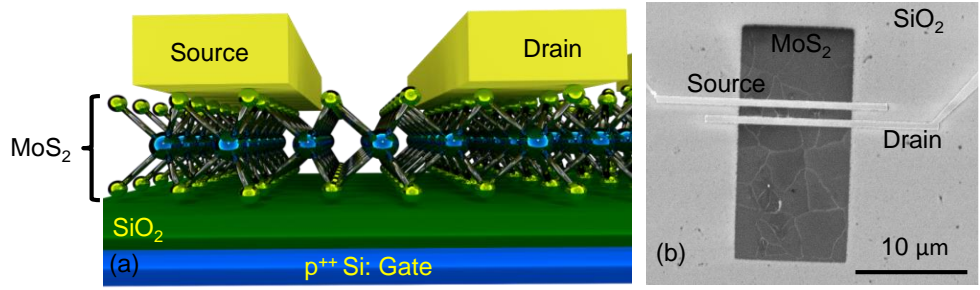

Figure 1: (a) Back-gated configuration of MoS$_2$ FETs with channel length, $L_{CH}$ = 1 µm and dielectric thickness, $T_{OX}$ = 90 nm. (b) Scanning electron microscope top-view of the fabricated FET. (Scalebar = 10 µm).

*Hysteresis induced resistance switching and its impact on transistor behavior*

RS phenomenon in various materials is typically confirmed through the drain-to-source current-voltage hysteresis behavior. Such a behavior is observed when the drain-to-source voltage ($V_{DS}$) is varied in a "dual-sweep" mode. This essentially means that the direction of $V_{DS}$ sweep is varied and a RS mechanism manifests as dual valued resistance (or current) for the same voltage. This is known as hysteresis and is observed in all the devices that exhibit RS[19]. For MoS$_2$ devices tested during this work, hysteresis is observed only at relatively large values of $V_{DS}$, as shown in Fig. 2(a). It is found that the voltage at which RS occurs is different at different gate voltages. For device in ON state ($V_{GS}$ = 60 V), RS occurs at $V_{DS}$ = 25 V whereas for open gate device and device in OFF state (-60 V) RS occurs at $V_{DS}$ = 20 V, Fig. 2(b). Moreover, RS in devices in gate open condition and OFF state is found to be abrupt at $V_{DS}$ = 20 V. This is not true for devices in the ON state and a gradual transition to a low resistance state is observed at $V_{DS}$ = 25 V. Such an observation is unique as it suggests that hysteresis induced RS depends on the dominant transport mechanism in the device. It is important to note that minor charge carrier transport during the OFF state and gate floating condition happens through mid-gap states and those which are close to the conduction band minimum (CBM), respectively, with marginal density as variable range hopping at room temperature[20, 21]. It is possible that an abrupt increase in the current during hysteresis may be a manifestation of increased density of states within the bandgap. This is further discussed later in the section. To determine to what extent RS has occurred, two terminal resistance of the device is measured before and after hysteresis. It is observed that, irrespective of the transport mechanism dominant during hysteresis, two terminal MoS$_2$ resistor, post hysteresis cycle, exhibits higher conductance than a virgin resistor, Fig. 2(b). However, the extent to which the device conductance increases does depend on the transport mechanism dominant at the time of hysteresis. A remarkable, 180% increase in the two terminal conductance is observed for the device subjected to hysteresis during OFF state operation. For device in the ON state and close to the threshold regime, increase in the conductance is found to be 80% and 55% respectively. These values are extracted from the $I_D V_D$ characteristic, shown in fig. S2 of SI, obtained for the two terminal MoS$_2$ resistor device. Further insights on the possible impact of hysteresis on MoS$_2$ are established by obtaining transfer characteristics of MoS$_2$ FETs before and after the hysteresis event. It is important to emphasize that transfer characteristics discussed next are of the same devices that are discussed in Fig. 2. Moreover, the current observed for hysteresis of device in the ON state (blue curve) is found to be comparable to that of a device in the OFF state which is not expected. It is important to note that devices are successively tested for hysteresis from a lower $V_{DS}$ to the point at which it exhibits hysteresis. It is found that the current degrades significantly due to successive attempts to observe hysteresis, especially when in the ON state, as shown in Fig. S2 of supplementary information (SI).

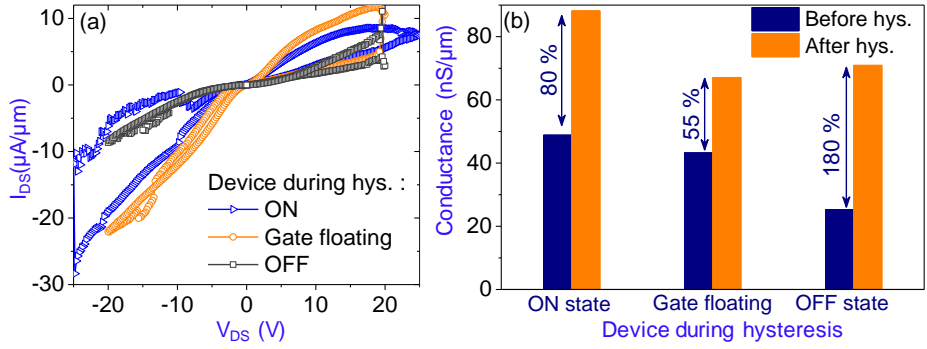

Figure *2*: (a) Hysteresis behavior observed when the transistor is in different operation regimes- ON state ($V_{GS}$ = 60 V), 2-terminal resistor (gate floating) and OFF state ($V_{GS}$ = -60 V). The drain-to-source voltage at which hysteresis occurs depends on which state the device is operated by tuning the gate bias. Hysteresis leads to significant change in material properties which manifests as change in channel conductance which is known as resistance switching (RS). (b) Hysteresis induced RS in our devices is found to increase the two-terminal resistance/conductance. In order to measure the 2-terminal conductance, $I_D$-$V_D$ characteristic is obtained while gate terminal is kept floating.

In general, it is found that devices after hysteresis exhibit degraded transistor behavior which is identified by huge negative shift in $V_T$, poor gate control and higher OFF state current, Fig. 3(a-c.) As shown in Fig. 3(a), hysteresis cycle at during the ON state ($V_{GS}$ = 60 V) results in complete loss of gate control and degraded ON state current. Whereas hysteresis at $V_{GS}$ = 0 and -60 V leads to significant negative shift in $V_T$, some loss in gate control and higher ON state current, shown in Fig. 3(b,c). From the transfer characteristics shown in Fig.3, it is clear that hysteresis-affected devices exhibit significant charge transport for the complete range of gate bias. As a result, negligible current modulation is observed. These results too suggest possibility of hysteresis induced increase in the density of mid-gap states which contribute to charge conduction when fermi-level is tuned below the conduction band minimum (CBM). It can be argued that hysteresis introduces incredibly large density of mid-gap states for which the device exhibits comparable currents in the otherwise distinguishable ON and OFF states. More importantly, this phenomenon occurs irrespective of the state in which the device operates i.e. independent of $V_{GS}$. Hence, this also explains the observed abruptness in the current hysteresis of devices in the OFF state and in two terminal configuration shown in Fig. 2(a). Prior to hysteresis, the density of mid-gap states is small in a high quality $MoS_2$ monolayer. As there are no defect peaks observed in the Raman spectra, shown in Fig. S1(b), it can be inferred that the defect density and hence density of mid-gap states is sufficiently low[22]. Post hysteresis, an extremely large increase in the density of mid-gap states contributing to the charge transport in the OFF state ($V_{GS}$ = -60 V) and near-band transport (floating gate) leads to an abrupt increase in the channel current. However, when the device is in ON state, increase in mid-gap states does not starkly affect charge transport because of the already pertaining transport through the conduction band with abundance of states available for charge transport. Since this is a post-hysteretic effect, it is believed to be related to the commonly observed hysteresis induced RS in $MoS_2$.

Interestingly, the effect of current-voltage hysteresis on transistor behavior is same as that of a steady-state/long-term DC electrical stress (LTES) as discussed in our previous work[17]. Such a remarkable similarity in the impact on transistor behavior and well-established hysteresis induced RS in $MoS_2$[9-15] encourages an investigation to validate the possibility of LTES-induced RS in our devices. Therefore, $MoS_2$ transistors are subjected to multiple LTES cycles with varying voltage polarity to observe change in the channel resistance.

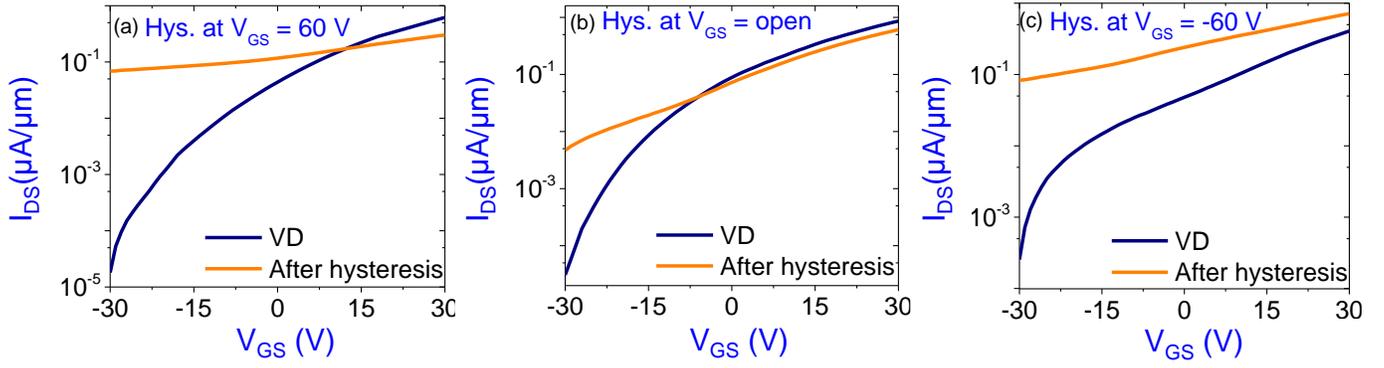

Figure 3: Transfer characteristics of MoS$_2$ FETs before and after the hysteresis cycle shown in figure 2(a). Hysteresis is performed on different devices which are operated in three major regimes of operation- (a) ON state, (b) 2-terminal resistor by keeping the gate terminal floating and (c) OFF state.

*LTES induced resistance switching*

In order to elucidate more on the effect of LTES on transistor behavior, MoS$_2$ devices are stressed as discussed in an earlier report[17] under conditions that prevent current decay during stress. Such stress conditions are chosen to ensure minimum lattice damage due to electron-phonon scattering (EPS) in order to de-couple RS mechanism from EPS induced change in material. A unique behavior is observed when a device is stressed at drain-to-source electric field, $E_{DS}$ = 0.2 MV/cm and $V_{GS}$ = 0 V for 300 seconds, Fig. 4(a). The current through the channel is found to be constant at $I_{DS}$ = 1µA/µm during the first 300 s stress cycle. Interestingly, the current starts-off from a higher value at $I_{DS}$ = 3 µA/µm in the second stress cycle and eventually rises abruptly to ~ 8 µA/µm at t = 200 sec and settles at ~ 4 µA/µm towards the end of the cycle at t = 300 sec. Current during subsequent stress cycles remains constant at ~ 4-5 µA/µm after which it again abruptly rises to ~ 11 µA/µm at t = 280 sec during the 5$^{th}$ stress cycle and the device continues to remain at the same current density throughout the 6$^{th}$ stress cycle. During the course of six stress cycles, the device exhibits three different resistance states. This is attributed to the electrical stress induced changes in MoS$_2$, earlier reported by the authors, where localized low resistance regions are found to have formed within the MoS$_2$ channel[17]. Effect of different cycles of stress induced material reconfiguration on the transistor behavior is shown in figure 4(b). On subsequent stress cycles, $V_T$ of the device is found to have shifted in the negative direction followed by a complete loss of gate control. This implies that the material has undergone changes such that the free-electron concentration in the channel has increased[17]. Different resistance states are observed in Fig. 4(a) which can be correlated to discrete increase in the electron concentration in the channel realized at different stress cycles. Observation of such discrete resistance states is similar to long-term potentiation (LTP) which is a synapse-like property of MoS$_2$ important for neuromorphic applications. In order to observe long-term depression (LTD) through LTES, devices are stressed with varying polarity of $V_{DS}$. Current variation during stress cycles of opposite polarity along with its impact on the transistor behavior is presented next.

Stress induced current fluctuations and perturbation of transistor behavior is found to be dependent on the polarity of drain-to-source bias. As shown in Fig. 5(a, b), alternate forward ($V_{DS}$ = 20 V) and reverse ($V_{DS}$ = -20 V) stress cycles result in abrupt or monotonically increasing and decreasing current respectively. A forward (reverse) stress brings the device in a low resistance state - LRS (high resistance state-HRS). As observed in fig. 5(c), device resistance fluctuates broadly between two values, within an error of ~25% in LRS and ~3% in HRS, when stressed alternately under opposite polarity bias, Fig. 5(d). In other words, positive stress leads to LTP whereas negative stress results in LTD. Moreover, the switching ratio, ratio between the resistance in LRS and that in HRS, is found to be ~ 2.76. This behavior is ascribed to resistance switching (RS) properties of MoS$_2$ wherein Sulfur vacancies migrate under the effect of electrical stress and a bipolar switching occurs, observed by others as discussed earlier[9-15].

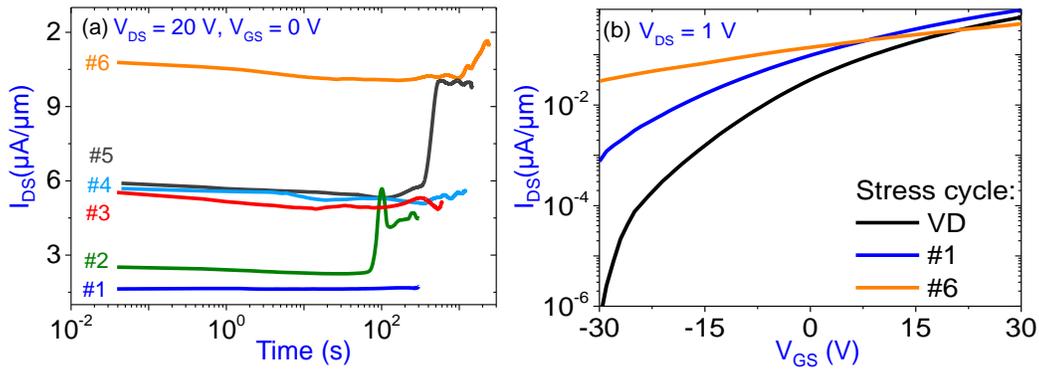

Figure 4: (a) Current variation during steady-state DC electrical stress as a function of number of stress cycles. Bias conditions during the stress are: $V_{DS}$ = 20 V and $V_{GS}$ = 0 V. (b) Impact of steady-state electrical stress on the transfer characteristics of $MoS_2$ FET. Perturbation from conventional transistor behavior is observed after long term stress and is found to be similar to that obtained after hysteresis in figure 3.

Such a variation with stress polarity is also observed in the transfer characteristics of the transistor, as shown in Fig. 5(e). Stress cycles with alternate polarity result in binary current fluctuation due to RS between the two resistance states. It is observed that, after the first positive stress cycle, gate tunability of the devices is significantly degraded along with a huge negative shift in $V_T$. Signature of LTD is found in transfer characteristics when a subsequent negative stress cycle tends to bring down the overall channel current. However, it fails to realize the original virgin state of $MoS_2$. This implies that, LTES results in a permanent change in $MoS_2$ which is irreversible in nature. Moreover, it also suggests that, akin current voltage hysteresis, discussed earlier, LTES triggers observable memory behavior in $MoS_2$. It is argued that, both hysteresis and LTES result in similar change in material configuration and hence perturbation in the behavior of $MoS_2$ transistors. Therefore, the above discussion suggests that RS in $MoS_2$ device can be triggered not only via hysteresis cycles but also via LTES with suitable stress conditions.

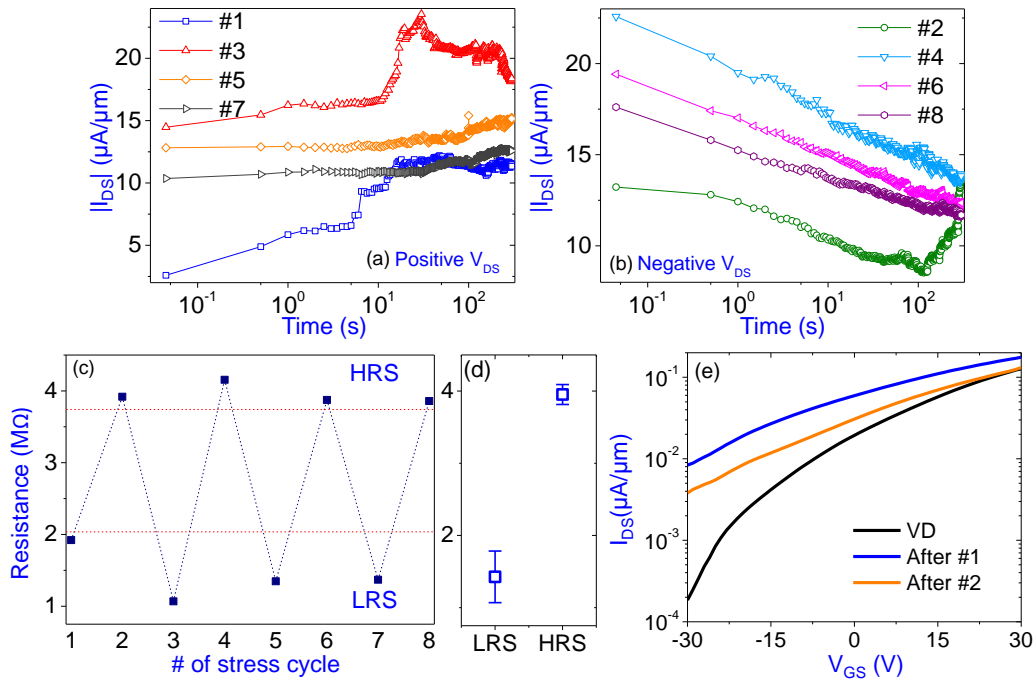

Figure 5: Current fluctuation during multiple cycles of steady-state electrical stress of (a) $V_{DS}$ = + 20 V and (b) $V_{DS}$ = - 20 V with $V_{GS}$ = 0 V in both cases. Note that the device is subjected to alternate positive and negative stress cycles (odd number cycles are positive while even number cycles are negative, shown in a and b respectively). (c) Two distinct resistance states are realized due to successive stress cycles. All positive/odd stress cycles led to a low resistance state (LRS) while all negative/even stress cycles resulted in a high resistance state (HRS). (d) The error/standard deviation in the values of two resistance states realized due to repeated operation is fairly low. (e) Effect of polarity of the steady-state stress on the transfer characteristics of $MoS_2$ FET. A negative stress following a positive stress event tends to nullify the effect of previous cycle on the transistor characteristics.

It can be inferred from Fig. 4 and 5 that LTES induces a permanent change in MoS$_2$ after which polarity dependent realization of two resistance states can be achieved. Similar behavior is observed in certain RS materials for which a preliminary forming process is used to trigger memory characteristics. This drives attention towards a possible forming step for MoS$_2$ which can improve its switching performance. It turns out that LTES induces better switching performance in a MoS$_2$ based two-terminal RRAM thereby making it more relevant for binary memory applications, as discussed next.

*Forming in MoS$_2$ for improved binary switching*

As discussed, LTES induced perturbation in MoS$_2$ is similar to voltage induced formation of RRAMs and other memory devices[19] reported earlier. Such a forming process has been reported to trigger memory behavior in HfO$_2$ based RRAMs[23] and is followed by SET/RESET programming of the memory cell. Moreover, the fact that LTES induces RS in our devices appeals to test this phenomenon as a possible forming process for CVD-grown monolayer MoS$_2$. In order to implement LTES as a forming process, 2-terminal MoS$_2$ devices are subjected to multiple SET and RESET cycles before and after LTES. This is done to observe the impact of LTES on binary switching behavior of these devices and address the question of a possible forming procedure in MoS$_2$.

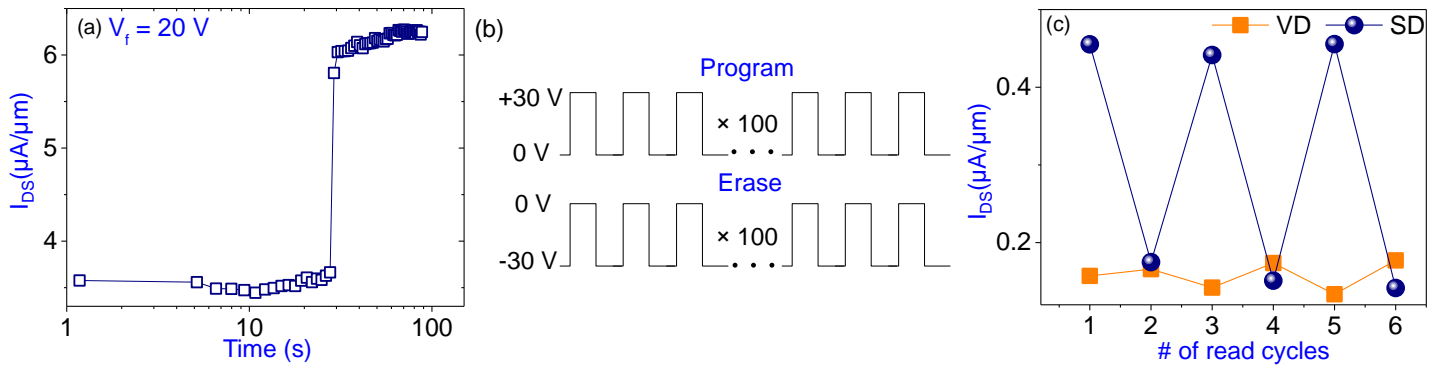

Figure 6: (a) Variation of device current during the forming process performed under the following stress conditions: $V_f$ or $V_{DS}$ = 20 V and $V_{GS}$ = 0 V. (b) Pulse program and erase signals are applied across the two terminal MoS$_2$ RRAM, before and after the electrical stress event shown in (a), to write '1' and '0' respectively. These signals are two trains each of 100 pulses of 100 ms and 50% duty cycle with amplitude of + 30 V and -30 V for programming and erasing respectively. (c) A read signal of 1ms width and 0.1 V amplitude is used to read the state of the device and is applied after every program or erase cycle. This is done successively for six cycles. As mentioned earlier, each program and erase cycle is followed by a read cycle and such a write-read sequence is used to analyze the switching behavior of MoS$_2$ before (VD) and after (SD) forming.

The forming event is carried out by stressing MoS$_2$ electrically at $V_{DS}$ = 20 V and $V_{GS}$ = 0 V for the duration until an abrupt change in the device current is observed. Such a scheme to form our devices is used because a sharp rise in current was earlier observed in Fig. 4(a) with the same bias conditions as mentioned above. It is observed in Fig. 6(a) that after a sustained current level for about 30 seconds, current rises sharply and settles at about 1.7 times higher value. Such a variation in current implies that the channel conductance has increased and therefore is regarded as a forming step. It is important to highlight the time instance at which current abruptly rises which is different in Fig. 6(a) than that shown earlier in Fig. 4(a). A possible argument for such a variation in device current dynamics is that grain boundaries, which are primarily responsible for RS in CVD-grown monolayer MoS$_2$[9], are randomly oriented and different devices can have different number of GBs running randomly across the channel region. In order to identify the effect of this forming step, switching performance of our devices is compared before and after forming. In order to perform a memory test, two terminal MoS$_2$ devices (open gate condition) are programmed (SET) and erased (RESET) by a train of 100 pulses of ± 30 V of 100 ms with 50% duty cycle, as shown in Fig. 6(b). In order to read the state of the device without perturbing it, a 1 ms read pulse of 0.1 V is applied across the device, Fig. 6(c). It is observed that, a virgin device (VD) does not exhibit significant binary switching. Multiple program and erase cycles result in consistent current (~ 180 nA/µm) which implies that the resistance of the device does not switch. This is contrary to the observed binary and analog switching in MoS$_2$[9-15]. Such a contrasting behavior is ascribed to program, erase and/or read bias conditions, insufficient to induce resistance change in the device. Identifying the right bias conditions for MoS$_2$ devices is not the focus of this work and is not discussed in the paper henceforth. It is important to emphasize that this paper discusses the effects of LTES on the switching behavior in MoS$_2$ devices and qualify it as a possible forming step. Interestingly and contrary to the observed failure in performing a SET/RESET process on a VD, the same device exhibits decent switching performance after undergoing a stress cycle/forming step and is referred as stressed device (SD). A train of 100 pulses of ± 30 V of 100 ms with 50% duty cycle to SET/RESET the device results in two different resistance states corresponding to two different current values, ~0.45 nA/μm (SET or 1) and ~0.18 nA/μm (RESET or 0), measured when a subsequent read pulse (0.1 V, 1 ms) is applied, as shown in Fig. 6(c). This is observed for six alternate SET and RESET cycles with a switching ratio (SET current/RESET current) of 2.5. These results clearly imply that LTES can be employed as a forming process for MoS$_2$ based two terminal RRAMs to ensure better switching performance.

As observed earlier in case of hysteresis induced conductance change in Fig. 2, gate terminal plays an important role during RS in our MoS$_2$ devices implying that transport mechanism in the MoS$_2$ determines the amount of change in channel resistance. We, therefore, explore the impact of gate bias on the switching performance, as obtained from pulse measurements scheme, of MoS$_2$. It is observed that irrespective of gate voltage, formed devices (FD) exhibit higher switching window (current at LRS – current at HRS) than virgin devices (VD), Fig. 7(a). Note that these devices are formed at $V_{DS}/V_F$ = 20 V and $V_{GS}$ = 0 V. The Switching ratio (SR) is found to have improved by ~29% at $V_{GS}$ = -30 V and the increase in SR is ~42% & ~11% at $V_{GS}$ = 0 V & 30 V respectively. This implies that SET/RESET cycles in both VD and FD are more efficient at lower and/or negative gate voltage and a forming process tends to improve the switching performance of VDs irrespective of the gate bias. Interestingly, the extent to which forming-led improvement in SR occurs depends on $V_{GS}$ and turns out that at $V_{GS}$ = -30 V and 0 V increase in SR is larger than that at $V_{GS}$ = 30 V. Note that while at $V_{GS}$ = 0 V, device is at the onset of ON state operation and charge transport happens via states close to CBM, at $V_{GS}$ = -30 V the device is in the OFF state due to transport through limited number of mid-gap states. Such a dependence on gate voltage is fallout of forming induced large density of mid-gap states that facilitate charge transport in MoS$_2$ in the OFF state, discussed earlier. A subtle yet important observation is that switching in MoS$_2$ is more efficient when device is in the OFF state even without a forming step, Fig. 7(a). These observations collectively imply that MoS$_2$ based RRAMs exhibit better switching performance when they are depleted of free charge carriers and a LTES forming step improves it further. It is worth mentioning that, owing to a large difference in the magnitude of current flowing in the three different regimes of operation, we choose SR as the parameter to compare switching performance at various gate voltages. While gate voltage has remarkable influence on RS properties of MoS$_2$, stress voltage during the forming step, $V_F$, also determines the switching performance of FDs. Devices formed at lower $V_F$ do not exhibit a significantly improved SW than VD. This is shown in the detailed transfer characteristics of VD and FD with varying $V_F$ in Fig. S3(a-d) of SI.

In order to observe the effect of $V_F$ on switching performance, MoS$_2$ devices are subjected to SET/RESET and read cycles before and after forming at varying $V_F$. The switching window of VDs and FDs are compared and shown in Fig. 7(b) as a function of $V_F$. Note that the gate terminal is kept open during these cycles for both VDs and FDs. This is done to ensure negligible effect of gate terminal on the switching behavior. The detailed $I_D V_D$ data for the gate open condition is shown in Figure S3 (e-h) of SI. It is observed that forming at higher $V_F$ leads to much higher improvement in the switching window (SW). SW is the difference between current at LRS and HRS. As shown in Figure 7(b), device formed at 20 V exhibits ~ 13× improved SW than its virgin counterpart. However, improvement in SW for devices formed at $V_F$ = 15 V, 10 V and 5 V is found to be ~ 1.7×, ~1.6× and ~1× respectively. Such a variation with $V_F$ suggests strong influence of the drain-to-source electric field during forming on RS properties of MoS$_2$. While $V_{GS}$ determines the dominant transport mechanism in MoS$_2$ FETs and lower or negative $V_{GS}$ facilitates better switching performance, it appears that $V_F$ triggers the formation of mid-gap states. Also, in order to introduce mid-gap states sufficient $V_F$ must be applied for an appropriate forming event to take place.

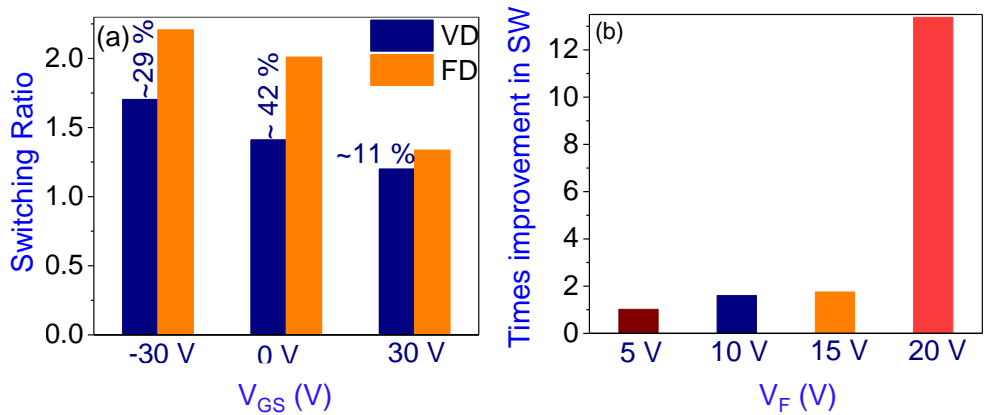

Figure 7: (a) Switching ratio (SR) as a function of gate voltage applied during program/erase and read cycles. SW is extracted from transfer characteristics shown in Fig. S3 of SI measured at $V_{DS}$ = 0.1 V. (b) SW as a function of forming voltage ($V_F$). Here SW is extracted for the 2-terminal device at $V_{DS}$ = 0.1 V via a voltage sweep event as shown in Fig. S3 of SI.

Insights developed on the impact of gate bias controlled charge transport and forming voltage on the RS properties of CVD-grown monolayer $MoS_2$ strongly suggest that RS is most efficient when the device is depleted of free charge carriers and the switching performance can be further improved by a high voltage forming step prior to use as binary memory cells in a circuit environment.

### III. Conclusion

We report new insights on a well-established phenomenon of current-voltage hysteresis which is known to induce resistance switching (RS) in CVD-grown monolayer $MoS_2$. The extent to which hysteresis affects resistance of the device is found to be dependent on the charge transport mechanism dominant in $MoS_2$ during hysteresis. Therefore, tuning gate bias to control the charge transport mechanism and hence RS in our devices is key for achieving improved switching performance. Impact of hysteresis on the overall transistor behavior suggests that RS in $MoS_2$ is accompanied by a unique deviation in the transistor behavior with poor gate control and huge negative shift in the threshold voltage. It turns out that such a deviation from the conventional transistor behavior is a signature of RS and is also observed when a long-term DC electrical stress (LTES) is applied on our devices. Polarity dependent realization of low resistance state (LRS) and high resistance state (HRS) confirm LTES induced RS. With this, besides hysteresis, we introduce another way of achieving RS in $MoS_2$. We also find that devices that undergo LTES exhibit significantly improved switching performance than virgin devices. Therefore, we propose that LTES, under certain bias conditions, can be used as a forming step in our devices to achieve enhanced memory switching window i.e. larger separation between LRS and HRS. RS induced by LTES is found to be linearly dependent on the drain-to-source voltage during stress, referred as VF, and therefore higher VF is found to induce enhanced switching window. Moreover, RS induced by hysteresis and LTES is more conspicuous for devices operating in the OFF state during which mid-gap states facilitate most of the charge transport across the channel via variable range hopping mechanism. Such gate voltage dependence of switching performance and impact of RS on transistor behavior clearly imply that both hysteresis and LTES introduce mid-gap states in $MoS_2$ that lead to RS. However, a more direct way of characterizing these states has not been discussed in this work. Developed insights on gate voltage dependence on RS and impact of VF on the switching window indicate that a higher forming voltage along with a negative gate bias to deplete the channel are essential to realize improved switching performance in CVD-grown monolayer $MoS_2$ based binary RRAMs and thereby making it more relevant for binary memory applications.

## Data Availability

The data to support findings of this work are available from the corresponding author upon reasonable request.

## Acknowledgment

Authors would like to thank NNetRA program of MeitY, DST and MHRD, Govt. of India, for supporting this work.

## Conflicts of interest

There are no conflicts to declare.

## Author Contribution

Ansh fabricated and characterized all the devices. Ansh, MS analyzed the data and wrote the paper.